\documentstyle[12pt,aasms4,flushrt]{article}

\def\myputfigure#1#2#3#4#5%
{\vskip#5pt\makebox[0pt]{\hskip#2in
\includegraphics[width=#3\textwidth]{#1}}\vskip#4pt\hfill}

\newcommand\lsim{\mathrel{\rlap{\lower4pt\hbox{\hskip1pt$\sim$}}
        \raise1pt\hbox{$<$}}}
\newcommand\gsim{\mathrel{\rlap{\lower4pt\hbox{\hskip1pt$\sim$}}
        \raise1pt\hbox{$>$}}}

\begin{document}
\Large 
\centerline{\bf Caught in the Act?}
\normalsize 
\author{\small Zolt\'{a}n Haiman}

\begin{verbatim}
                   News and Views, Nature, 26 August 2004
\end{verbatim}

\vskip 0.2in 

\hrule 

\vskip 0.1in 

\noindent
{\em\large Which came first, the stars and gas that make up a galaxy,
or the giant black hole at its centre? Observations of a distant
galaxy, caught as it forms, could help solve this chicken--and--egg
problem.}

\vskip 0.2in 

\hrule 

\vskip 0.2in 

\thispagestyle{empty}

Galaxies are thought to be surrounded by massive halos of dark matter,
each outweighing its galaxy by a factor of about eight. The visible
part of a galaxy, occupying the inner 10\% of the halo, consists of a
mixture of stars and gas.  Galaxies harbour a giant black hole at
their centres, which in some cases is actively fuelled as it sucks in
surrounding gas. In especially active galaxies, called quasars, the
fuelling rate is so high that the radiation generated close to the
black hole outshines the cumulative star--light from the entire
galaxy. The sequence of cosmic events that leads to this configuration
is still largely mysterious. How does gas condense into the central
regions of the dark--matter halo? At what stage of the gas condensation
process do the stars and the giant black hole light up?  The unique
observation of a distant quasar by Weidinger {\it et al}.\cite{weidinger},
reported in this issue, offers fresh insight.

The formation of massive dark--matter halos is determined by gravity,
and can be described by using ab initio calculations\cite{nfw96}. As
the Universe expanded from its dense beginning, tiny inhomogeneities
in the distribution of dark matter were amplified through the effects
of gravity. Regions of space that were slightly denser than average
had a higher gravitational pull on their surroundings; eventually
these regions stopped following the expansion of the rest of the
Universe, turned around and re--collapsed on themselves. The resulting
dense knots of dark matter -- forming the intersections of a cosmic
web of more mildly overdense dark--matter filaments -- are believed to
be the sites at which galaxies lit up.

Dark matter thus dominates the formation of a galaxy, at least
initially, and determines the gross properties of the galaxy
population, such as their abundance, size and spatial
distribution. But it is the trace amount of gas (mostly hydrogen and
helium), pulled with the dark matter into the collapsed halos, that
forms the visible parts of galaxies and determines their observable
properties. In particular, to condense to the core of the dark halo
the gas has to cool continuously, to deflate the pressure acquired by
its compression. A fraction of the gas (typically 10\% by mass)
eventually turns into stars, and a much smaller fraction (typically
0.1\%) into the central massive black hole\cite{magoo}.

The composition of the gas inside the galaxy can be studied through
the spectrum of radiation it emits and absorbs.  Primordial gas is
essentially a pure mix of hydrogen and helium, but all of the quasars
discovered so far have shown the presence of various heavier elements
(such as carbon, nitrogen, oxygen and iron). This indicates that the
gas has been enriched by the nucleosynthetic yields from previous
generations of stars. Even the most distant quasars, including those
that existed about 1 billion years after the Big Bang (a mere 5\% of
the current age of the Universe), show significant heavy--element
content\cite{fan}. This suggests that vigorous star--formation is a
necessary condition for any quasar activity. On the other hand, star
formation seems to be confined to relatively small scales, close to
the galactic centre. A natural inference would then be the following
sequence of events: the cosmic gas first contracts to the inner
regions of the halo and only then forms stars -- but this is still
before the formation (or at least activation) of any central quasar
black hole.

Not necessarily so, according to Weidinger {\it et al.}\cite{weidinger}, who
have detected the faint glow of hydrogen emission enveloping a distant
quasar at a radius equivalent to about 100,000 light years -- several
times the size of the visible part of a typical galaxy. Such emission
has a simple physical origin. The hydrogen atoms falling through the
halo are ionized by the quasar's light, then recombine with electrons
to become atoms again. Each recombination results in the emission of a
so--called Lyman--$\alpha$ photon (a photon with energy equal to the
difference between the ground and first excited state of a hydrogen
atom). As a result, when viewed through a filter tuned to the
Lyman--$\alpha$ frequency, a faint 'fuzz' can be seen to surround
quasars\cite{rees88}. This fuzz can serve as a diagnostic of whether
or not a spatially extended distribution of infalling gas is present
around the quasar\cite{hr01}. If most of the gas has already cooled
and settled at the centre of the halo, the extended fuzz would be
absent.

Although the fuzz detected by Weidinger {\it et al.} is faint, it is as
bright as would be expected if all of the hydrogen needed to make up a
typical galaxy is still infalling\cite{hr01}. As a result, the galaxy
imaged by Weidinger {\it et al.} is likely to be still in its infancy,
despite the fully--formed appearance of its bright central quasar black
hole. From the shape and kinematics of the fuzz, the authors were also
able to confirm the presence of the dark--matter halo, which is
accelerating the infall of the hydrogen gas, and to measure the halo's
mass. Their value -- $2-7 \times 10^{12}$ solar masses -- is in
accord with independent estimates from spectral absorption
features\cite{bl03} and from the abundance\cite{hh01} of other quasars
of similar brightness.

Extended Lyman--$\alpha$ emission has previously been detected around
quasars, but its interpretation as being due to a wide region of
infalling gas was less convincing\cite{m00}$^-$\cite{bergeron}. In
contrast, Weidinger {\it et al.}\cite{weidinger} have obtained a
two--dimensional image of the source clearly, showing the extent of
the emission, and a spectrum showing a velocity pattern consistent
with infall.  Extended emission is a relatively common phenomenon
among so--called radio--loud quasars -- a special subset believed to
drive powerful gaseous outflows. However, the gas surrounding these
sources is always enriched with heavy elements, whereas the fuzz
detected the Weidinger {\it et al.} appears to be pristine (no
heavy--element absorption or emission features are detected).

This result will undoubtedly prompt further theoretical modelling of
the density and velocity distribution of the infalling gas.  An
outstanding question, for example, is whether the gas could have been
delivered by a recent merger with another galaxy; such mergers stir up
and spread gas over large regions, and might also be responsible for
having activated the quasar in the first place.  An observational
census of similar quasars should be equally interesting, and feasible
with a modest investment of time on modern telescopes. It could show
whether prompt activation of a quasar, at an early stage in the
assembly of the galaxy while there is still a significant amount of
cosmic hydrogen gas--infall, is the norm among galaxies, or whether it
is a peculiarity of the galaxy examined by Weidinger and colleagues.
In the meantime, we have a new piece in the chicken--and--egg puzzle,
suggesting that, at least occasionally, the egg -- the black hole --
may come first. $\Box$

\small

\noindent{\it Zolt\'an Haiman is in the Department of Astronomy,\\ Columbia
 University, New York, NY 10027, USA.\\email: zoltan@astro.columbia.edu}


\begin{thebibliography}{}

\bibitem[$^1$]{weidinger}\cite{weidinger} Weidinger, M., M{\o}ller, P. \& Fynbo, J. P. U. {\it Nature}, {\bf 427}, xxx--yyy (2004).
\bibitem[$^2$]{nfw96}\cite{nfw96} Navarro, J. F., Frenk, C. S., \& White, S. D. M. {\it Astrophys. J.}, {\bf 462} 563--575 (1996).
\bibitem[$^3$]{magoo}\cite{magoo} Magorrian, J., {\it et al.} {\it Astron. J.}, {\bf 115}, 2285--2305 (1998).
\bibitem[$^4$]{fan}\cite{fan} Fan, X. {\it et al.} {\it Astron J.}, in press, astro-ph/0405138 (2004).
\bibitem[$^5$]{rees88}\cite{rees88} Rees, M. J. {\it Mon. Not. Roy. Ast. Soc.}, {\bf 231}, 91p--95p (1988).
\bibitem[$^6$]{hr01}\cite{hr01} Haiman, Z., \& Rees, M. J. {\it Astrophys. J.}, {\bf 556}, 87--92 (2001).
\bibitem[$^7$]{bl03}\cite{bl03} Barkana, R., \& Loeb, A. {\it Nature}, {\bf 421}, 341--343 (2003).
\bibitem[$^8$]{hh01}\cite{hh01} Haiman, Z., \& Hui, L. {\it Astrophys. J.}, {\bf 547}, 27--38 (2001).
\bibitem[$^9$]{m00}\cite{m00} M{\o}ller, P., Warren, S. J., Fall, S. M., Jakobsen, P., \& Fynbo, J. U. {\it The ESO Messenger}, {\bf
99}, 33--35 (2000).
\bibitem[$^{10}$]{bunker}\cite{bunker} Bunker, A., Smith, J., Spinrad, H.,  Stern, D., \& Warren, S. {\it Astroph. Sp. Sci.}, {\bf 284}, 357--360 (2003).
\bibitem[$^{11}$]{bergeron}\cite{bergeron} Bergeron, J., Petitjean, P., Cristiani, S., Arnouts, S., Bresolin, F., \& Fasano, G. {\it Astron. Astrophys.}, {\bf 343}, L40--44 (1999).

\end{thebibliography}
\end{document}